# Pulsed Accretion in a Variable Protostar

James Muzerolle[1], Elise Furlan[2], Kevin Flaherty[3], Zoltan Balog[4], Robert Gutermuth[5]

[1] Space Telescope Science Institute, 3700 San Martin Dr., Baltimore, MD 21218, USA
[2] National Optical Astronomy Observatory, Tucson, AZ, 85719, USA
[3] Steward Observatory, 933 N. Cherry Ave., The University of Arizona, Tucson, AZ 85721, USA
[4] Max-Planck-Institut fur Astronomie, Konigstuhl 17, 69117, Heidelberg, Germany
[5] Department of Astronomy, University of Massachusetts, Amherst, MA 01003, USA

**Periodic increases in luminosity arising from variable accretion rates have been predicted for some close pre-main sequence binary stars as they grow from circumbinary disks[1-3]. The phenomenon is known as "pulsed accretion" and can affect the orbital evolution and mass distribution of young binaries[2,4], as well as the potential for planet formation in the circumbinary environment[5,6]. Accretion variability is a common feature of young stars, with a large range of amplitudes and timescales as measured from multi-epoch observations at optical[7,8] and infrared[9-13] wavelengths. Periodic variations consistent with pulsed accretion have been seen in only a few young binaries via optical accretion tracers[14-16], albeit intermittently with accretion luminosity variations ranging from 0-50 percent from orbit to orbit. Here we report on a young protostar (age ~$10^5$ yr) that exhibits periodic variability in which the infrared luminosity increases by a factor of 10 in roughly one week every 25.34 days. We attribute this to pulsed accretion associated with an unseen binary companion. The strength and regularity of this accretion signal is surprising; it may be related to the very young age of the system, which is a factor of ten younger than the other pulsed accretors previously studied.**

We obtained multi-epoch mid-infrared (MIR) observations of the star forming region IC 348 using the Spitzer Space Telescope. Among the roughly 300 pre-main sequence objects in the cluster, the protostar LRLL 54361 (hereafter, L54361) exhibits by far the largest MIR flux variability. We have a total of 81 separate observations of L54361 taken with all three instruments on board Spitzer. The multi-epoch spectral energy distribution (SED) is shown in Figure 1. The measured bolometric luminosity of the system ranges from about 0.2 to 2.7 $L_\odot$. The spectral shape remains relatively constant over this range, aside from slightly bluer mid-infrared colors at higher luminosities.

The photometric light curve indicates that the flux variations occur repeatedly throughout the 7-year span of our observations. The two longest contiguous sets of photometry (Figure 2) reveal a strong pulse signature in which the flux increases by about 2 magnitudes in as little as a few days, followed by a longer exponential decay over the following few weeks. The combined photometric dataset suggests that the variability of L54361 is periodic in nature; the pulse shape revealed by the contiguous warm Spitzer photometry appears in the older data and at other wavelengths, albeit with variations in the pulse width and peak flux. Using several statistical tests, we find that the flux peaks repeat with a robust period of 25.34 ± 0.01 days.

Follow-up multi-epoch imaging taken with the Hubble Space Telescope at near infrared wavelengths reveals spatially-resolved scattered light structures associated with L54361 (Figure 3). The central source varies with almost the same amplitude and light curve shape at 1.6 microns as seen in the Spitzer data, and the peak occurs exactly as expected given the previously determined period. The geometry of the scattered light is similar to that of other protostars[17], and is likely produced by cavities carved out of an infalling envelope by one or more outflows. The apparent motion of the scattered light indicates a light echo produced as the pulse peak light travels through the outflow cavities, and suggests that the source of the illumination is relatively isotropic.

There are three primary sources of periodicity in YSOs: stellar rotation, Keplerian rotation of an inner disk, and the orbital motion of a close binary companion. Stellar rotation can manifest itself via localized hot or cool spots, or via interactions between the stellar magnetic field and the inner disk. In the case of L54361, we reject stellar rotation effects on several grounds: 1) rotation periods for pre-main sequence stars range from a few days to two weeks[18] (and protostars are typically faster still[19]), all lower than the measured period; 2) dark spots produce sinusoidal light curves, with amplitudes of a few tenths of a magnitude in the optical and declining to longer wavelengths; 3) hot spots tend to produce less obvious periodicity owing to the more stochastic nature of accretion, and can exhibit phase-dependent asymmetric illumination of the circumstellar material as they rotate with the star[20], which we do not see.

Regarding phenomena related to Keplerian rotation of an inner disk, persistent asymmetric structures such as warps in the inner disk can produce periodic obscuration of both single stars[7] and binary systems[21,22]. We argue that the data are not consistent with this scenario in several respects: 1) obscuration localized to the disk plane would not affect light propagating in the perpendicular direction through the outflow cavity; 2) obscuration events produce characteristic light curve "dips", while we see a positive pulse-like shape; 3) as we show below, the MIR and far-infrared flux of L54361 originates mostly in the infalling envelope, whose total flux would not be significantly affected by localized stellar obscuration.

The third possibility, a connection to binary motion, is plausible in terms of the length of the period of L54361, although we do not yet have direct evidence of a companion. The pulsed accretion scenario could explain both the light curve shape and amplitude. Circumbinary disk simulations consistently show gap-clearing by gravitational torques, followed by accretion streams that feed material onto the central stars[1-3]. For certain binary architectures, particularly in the case of a highly eccentric orbit, the stellar accretion depends on orbital phase, with the highest accretion rates typically associated with periastron passages. A qualitatively similar process has also been suggested for some X-ray binaries, at least one of which has exhibited optical light curves similar to the MIR behavior of L54361[23,24].

An increase in the accretion luminosity as a result of the binary interaction increases the irradiation heating of circumstellar dust, which then reradiates the energy in the MIR where we observe it. To help test this hypothesis against our observations, we calculated radiative transfer models of protostellar dust emission and scattering[25]. The models include three components that are typical of protostellar systems: infalling envelope, accretion disk, and central star. Holding all parameters fixed except for the accretion luminosity, we are able to match the change between SEDs corresponding to two different pulse phases (Figure 4). The models show a relatively weak wavelength dependence as a function of luminosity, with a slightly flatter spectral slope at about 15 to 70 microns at higher luminosity as a result of optical depth effects in the envelope, in relatively good agreement with the observations.

We do not yet have any direct measure of the central object or its multiplicity status. As our models show, however, the bolometric luminosity provides an estimate of the stellar plus accretion luminosity. Assuming that the low end of the measured range is representative of the stellar luminosity, the combined stellar mass can be roughly estimated by comparing to a theoretical protostellar birthline[26] on an Hertzsprung-Russell diagram. We derive a value of ~ 0.2 $M_\odot$ (probably an upper limit because some contribution from accretion is likely, although the luminosity may also be somewhat underestimated because of scattering). Conversely, assuming that the upper end of the range of measured luminosity is due entirely to accretion luminosity and adopting the above stellar mass, we derive a maximum mass accretion rate of $10^{-6}$ $M_\odot$ yr$^{-1}$. This is at the upper end of the range of values measured from standard accretion diagnostics[27]. Spectroscopic observations are needed to verify an accretion signature, as well as characterize the binary orbit.

Why L54361 exhibits such a strong and regular signature, unlike the T Tauri-type pulsed accretors observed previously, remains unknown. There may be a connection to its earlier evolutionary stage, in which the infalling envelope provides a steady supply of material to the circumbinary disk. By contrast, T Tauri binaries are older by about a factor of ten, have long since dissipated their natal envelopes, and accrete at lower mean rates. Perhaps stochastic variability from other sources such as stellar magnetic interactions or disk turbulence can overwhelm the periodic signature in older stars. It is also possible that the particular orbital parameters of L54361 are rare but more favorable for modulating the accretion flow, such as a very large eccentricity.

**Acknowledgements**
This work was supported in part by NASA through Spitzer and HST GO contracts.  We thank S. Lubow, M. Livio, and N. Calvet for discussions.  E. F. was visiting the Infrared Processing and Analysis Center, Caltech, during the course of this work.

**Author Contributions**
J. M. and K. F. designed the Spitzer observations.  Z. B. and R. G. reduced the IRAC images and compiled the photometry, while J. M. reduced and analyzed the MIPS data.  J. M. and E. F. extracted and analyzed the IRS spectroscopy.  J. M. designed the HST observations and analyzed the images.  E. F. calculated the radiative transfer models and fit the observed SEDs.  All authors contributed to the writing of the paper.

**Author Information**
The authors declare no competing financial interests.  Correspondence and requests for materials should be addressed to J. M. (muzerol@stsci.edu).

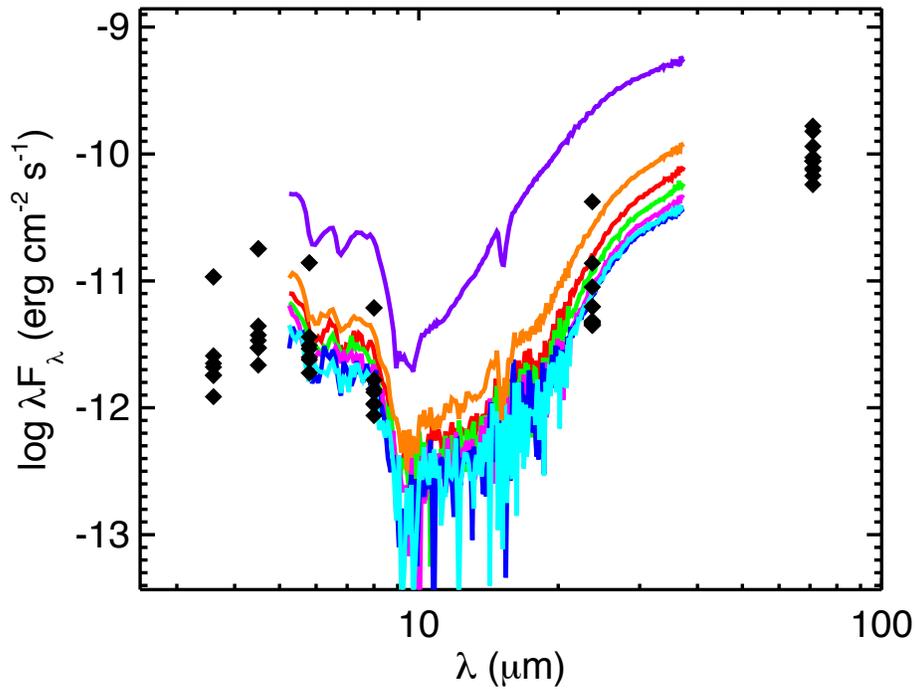

Figure 1. Multi-epoch spectral energy distribution of L54361. Our complete set of observations taken during cryogenic Spitzer operations are shown, including photometry from all four IRAC channels at 3.6 - 8 microns and the MIPS 24 and 70 micron channels (diamonds), as well as 7 epochs of IRS spectroscopy (in chronological order: red, green, magenta, blue, cyan, purple, and orange lines). Each single-epoch SED exhibits a shape characteristic of Class I objects, with the flux rising sharply to longer wavelengths and a strong silicate absorption feature at 8-12 microns. Between epochs, however, the flux varies by as much as an order of magnitude at all wavelengths, with only a slightly shallower spectral slope in the ~15-70 micron continuum as the flux increases.

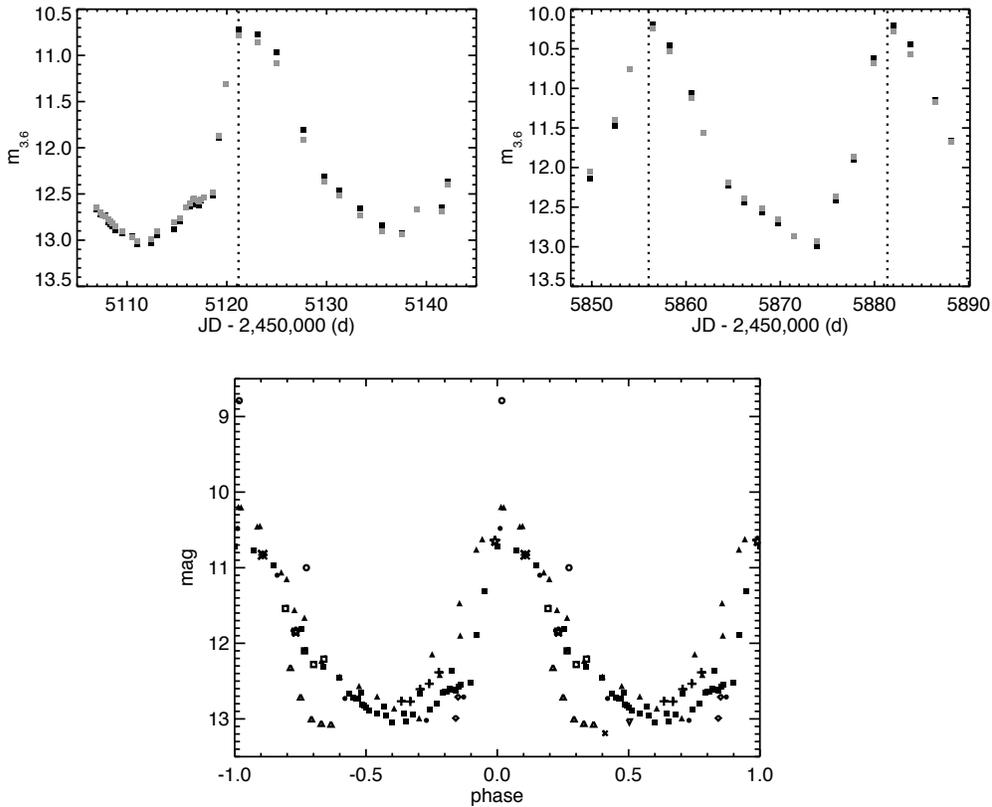

Figure 2. Photometric light curves for L54361. *Upper panels:* IRAC 3.6 (black) and 4.5 (gray, scaled down to match 3.6) micron magnitudes from the fall 2009 (left) and fall 2011 (right) Spitzer observing campaigns. Note that the 3-sigma photometric uncertainties are equal to or smaller than the symbol size. The dashed line in the left panel marks the observed peak time, which we set as the fiducial epoch for phase=0. The dashed lines in the right panel mark the predicted peak times assuming a periodicity of 25.34 days. *Lower panel:* The phased photometric light curve of L54361, assuming a period of 25.34 days and the phase zero epoch JD 2,455,121.203. Included are measurements taken at three separate wavelengths. Each symbol type represents a contiguous set of photometry: cryo-Spitzer IRAC 3.6 microns (plus signs, asterisk, cross), warm Spitzer IRAC 3.6 microns (solid squares, solid triangles), MIPS 24 microns (inverted triangle, open diamonds, open stars, open triangles), IRS 24 microns (open circles, open squares), HST WFC3 1.6 microns (solid circles). The 24 micron photometry are offset by +7.3 magnitudes and WFC3 photometry offset by -5.3 magnitudes in order to place everything on the same scale.

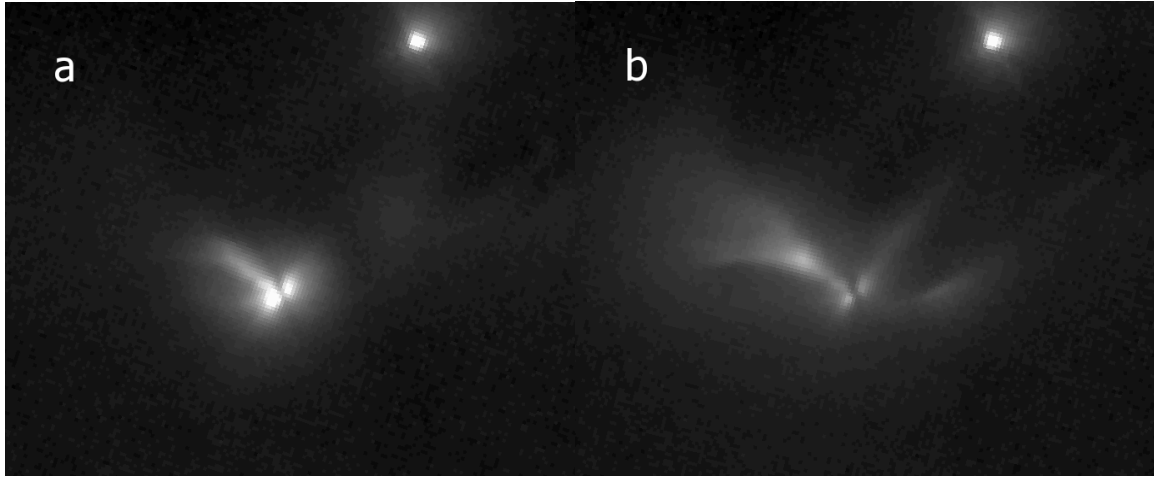

Figure 3. Near-infrared images of L54361. The two panels show portions of images taken with HST/WFC3 at 1.6 microns at two epochs corresponding to pulse phases of 0 (a) and 0.3 (b). North is up and east is to the left. L54361 is the extended source just below the center of the images; the point source at upper right is another YSO, LRLL 1843. The light from L54361 subtends roughly 14" (~4000 AU at the distance of the IC 348 region) in (a), and about 50" (~15,000 AU) in (b). Most if not all of this light is likely the result of scattering off of circumstellar dust in the protostellar envelope. An apparent edge-on disk is visible at the center of the object, and 3 separate structures indicative of outflow cavities extend to the northwest, southwest, and northeast. The extent and morphology of the scattered light changes substantially between epochs as a result of the propagation of the pulse peak light. (See Supplementary Information for the complete set of HST images.)

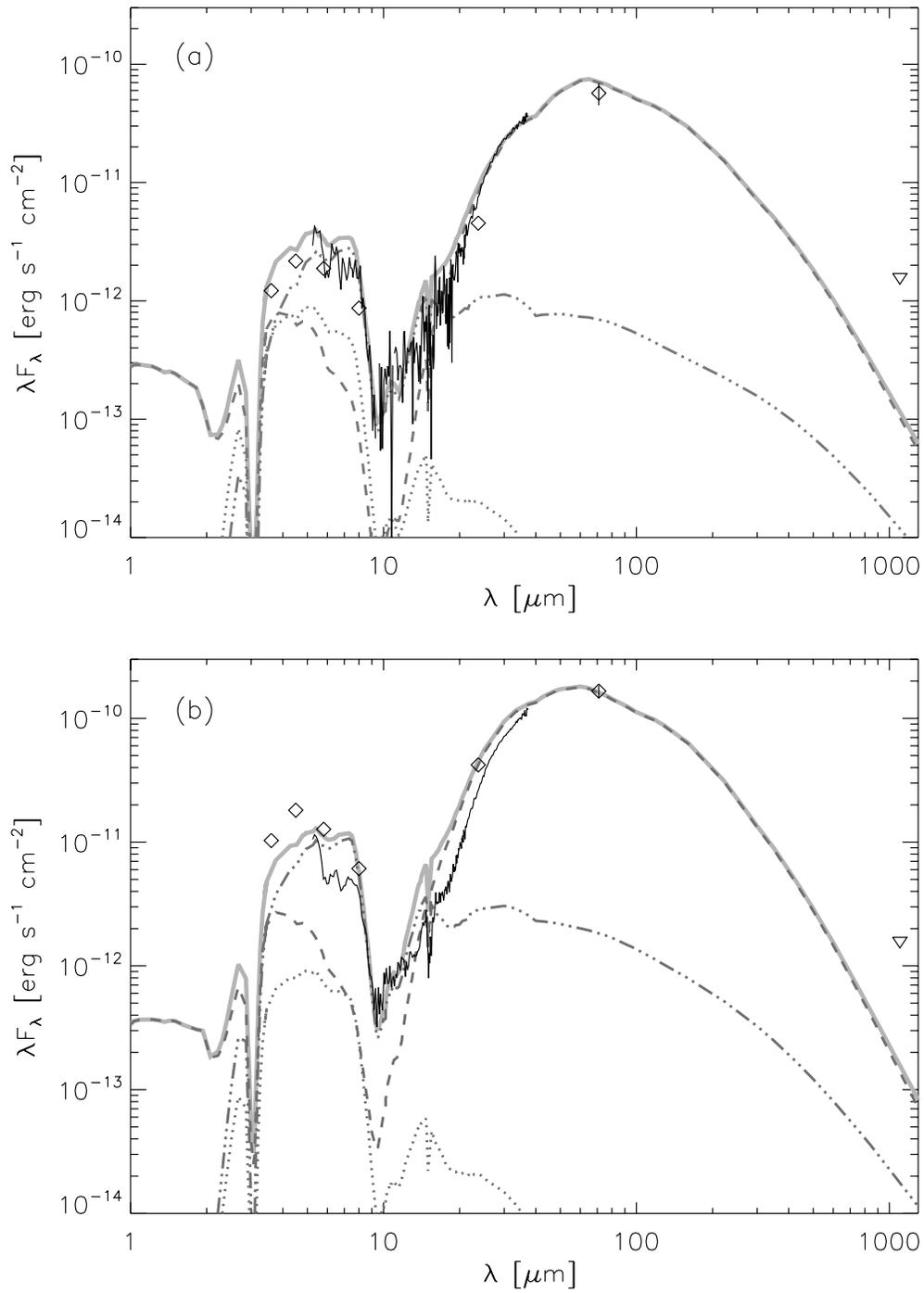

Figure 4. Protostellar spectral energy distribution models for L54361. We compare models to two sets of observations roughly representing pulse phases of 0.4 and 0.15 (top and bottom panel, respectively). The photometry and spectroscopy are not simultaneous but were selected to correspond roughly to a common flux level. The dashed lines indicate scattering and emission from the infalling envelope; the dash-triple-dotted lines represent scattering and emission from the circumstellar accretion disk; the dotted lines represent reddened flux from the central star; the solid gray line shows the total flux from all these components. The best-fit parameters include an envelope infall rate of $3 \times 10^{-6}$ $M_\odot$ yr$^{-1}$ (assuming a stellar mass of 0.2 $M_\odot$), envelope centrifugal radius $R_c = 30$ AU, outflow cavity opening angle $\theta = 30°$, inclination angle of the outflow/stellar rotation axis to the line of sight $i = 70°$, total central luminosity (stellar plus accretion) $L = 0.5$ $L_\odot$ (top panel) and 1.3 $L_\odot$ (bottom panel), and the fraction of $L$ due to the stellar luminosity $\eta_{star} = 0.5$ (top panel) and 0.19 (bottom panel). The only parameter that was actually changed between the two models was $\eta_{star}$, with appropriate values so that the stellar luminosity remained constant while the accretion luminosity changed by a factor of ~4. (See Supplemental Information for details of the model calculations and parameters.)

## S1 Supplementary Information

**Spitzer Observations & Data Reduction**

We observed L54361 with all instruments onboard the Spitzer Space Telescope; Table S1 shows the observation log. IRAC observations obtained during the cryogenic mission comprise all four channels (3.6, 4.5, 5.8, and 8 microns), while the warm operations data comprise only the first two. All frames were processed using the Spitzer Science Center (SSC) IRAC Pipeline v14.0.0, and mosaics were created from the basic calibrated data (BCD) frames using a custom IDL program[28]. Aperture photometry on these images was carried out using PhotVis version 1.10, which is an IDL GUI-based photometry visualization tool. The relative photometric uncertainties are < 1%. The zero point magnitudes of the calibration were 19.6642, 18.9276, 16.8468, and 17.3909 for channel 1, 2, 3, and 4, respectively. Aperture corrections of 0.21, 0.23, 0.35 and 0.5 mag were applied for channels 1, 2, 3, and 4 to account for the differences between the aperture sizes used for the standard stars and for the IC 348 photometry. We also reprocessed existing archived data, from GTO (PID 6 and 58) and c2d legacy (PID 178) programs, using the same methodology. The photometry is shown in Table S2.

For spectroscopy with the IRS, both Short-Low and Long-Low modules were used in order to cover the full wavelength range, 5 to 40 microns, with spectral resolution $\lambda/\Delta\lambda \sim 60\text{-}120$. The observations were performed in staring mode using two cycles of 14 seconds exposure for both modules. The data were reduced using SMART version 8.1.2[29], starting with the BCD products from the SSC reduction pipeline version S18.7. The two-dimensional spectra taken at two nod positions were subtracted from each other in order to remove the background. The spectra were then extracted using optimal source extraction[30]. The residual local sky background (generally very low) was subtracted by fitting a polynomial to the sky pixels on either side of the source at each row in the spatial direction. The final calibrated spectra for each nod and order were then combined using the sigma-clipped averaging function in SMART. Intrinsic measurement uncertainties are about 2% at the highest flux levels, though the S/N is much lower at fainter levels particularly in the region of the 10 micron silicate absorption feature.

The MIPS observations were taken in scan mode using 12 scan legs with 30' length, half-array offsets, and medium scan rate, with coordinates centered on the IC 348 cluster center. Starting with the raw data, individual images were calibrated and mosaicked using the MIPS GTO team Data Analysis Tool version 3.06. Photometry was measured using the daophot PSF fitting routine. We only analyzed the 24 and 70 micron data, as the 160 micron channel was saturated. We also reprocessed existing archived data, from GTO (PID 6 and 58) and c2d legacy (PID 178) programs, using the same methodology. In all images, the relative photometric uncertainties are <1% at 24 microns and ~10% at 70 microns. The photometry is listed in Table S3.

**Periodicity Analysis**

We considered all Spitzer photometry of L54361 obtained with the IRAC 3.6 micron and MIPS 24 micron channels. In addition, we added 24 micron photometry derived from the IRS data by convolving the spectra with the MIPS filter throughput. A Lomb-Scargle periodogram analysis of this entire set of photometry shows a significant peak at a period of 25.34 days (Figure S1). This period has a false alarm probability of $2 \times 10^{-8}$. As a check, we also ran the same data with the "SigSpec" routine[31], which is statistically more robust than the Lomb-Scargle periodogram. SigSpec returns a probability of $1.3 \times 10^{-11}$ that the 25.34-day period is not real. We conservatively estimate an uncertainty on the period determination of 0.01 days, based on the FWHM of the periodogram peak; a formal analysis using traditional periodogram methodology[31] yields an uncertainty of 0.0007 days.

**HST Observations and Analysis**

We obtained 7 epochs of imaging with the Hubble Space Telescope WFC3/IR instrument, using the F160W broad band filter. A 4-point dither pattern was used, with exposure times of about 10 minutes per point. The data were processed using the STScI pipeline, including MultiDrizzle combination of individual exposures and subtraction of the mean background level. For comparison with the Spitzer photometry, as shown in Figure 2 of the main text, we measured aperture photometry of the central source using an aperture radius of 1" centered at the position of the MIR source (this includes both scattered light lobes on either side of the apparent edge-on disk) and no background subtraction. Instrumental magnitudes were converted to the Vega system using a zeropoint magnitude of 24.7.

Portions of each image showing the full visible extent of L54361 are shown in Figure S2. The central source is split into two lobes separated by a dark lane indicative of an edge-on disk; it is probably not seen directly but rather via scattered light on both sides of the disk. The extended nebulosity includes three V-shaped structures extending to the northwest, northeast, and southeast (the latter considerably foreshortened), with their vertices meeting approximately at the central position of the MIR source (Figure S3). The morphology of these regions is typical of protostellar sources, and is the likely result of cavities carved out of the protostellar envelope by one or more outflows. Knots of emission likely related to a jet/outflow, most of them seen previously from the ground via molecular hydrogen emission[33,34], extend further to the northwest. The different angle of the NE structure suggests either that the outflow opening angle is larger on the east side of the source, or that there is a separate jet/outflow (though one has not been detected directly thus far). There is also a very faint feature in some of the images that appears to connect the NE cavity to the nearby object LRLL 1843. This source is a known substellar member of the cluster with a spectral type of M8.75, mass ~0.02 $M_\odot$, age < 1 Myr[35]. Whether the two objects are physically related cannot be determined with the current data.

The brightness and physical extent of the scattered light are highly variable, with the nebulosity growing larger in size after the central source reaches its peak flux. These morphological variations strongly suggest the presence of a light echo. Moreover, they also indicate that the illumination resulting from the flux pulses is more or less isotropic, rather than localized as would be expected for a rotating stellar hot spot or an obscuration event associated with the inner disk. We measured photometry at various small apertures spaced with increasing distance from the central source along two different axes roughly corresponding to the NW/SE and NE outflow cavities (Fig. S4). The resultant relative magnitudes are shown for each aperture as a function of observation epoch in Figure S5. The shape and relative amplitude of the pulse are clearly preserved in all apertures. The timing of the peak flux is shifted to consistently later times as one goes farther from the central source, as expected for a light echo. The lag between the arrival time of light observed from an echo at a projected distance *r* in arcseconds from the source and light observed directly from the source can be expressed as

$$\Delta t = \frac{rd}{c}\frac{1-\cos\theta}{\sin\theta}, \quad (1)$$

where $\theta$ is the angle of the light echo location relative to the line of sight, and *d* is the distance to the central source in parsecs[36]. Using the folded Spitzer IRAC photometry as a proxy for the intrinsic pulse profile from the central source, we can reasonably match most of the HST aperture light curves by applying the appropriate delay times as given by Equation 1, assuming an angle $\theta = 90°$ (i.e., in the plane of the sky) and a distance $d = 250$ pc. The fits to apertures E and F could be improved by adopting $\theta = 70°$; this may indicate that the NE structure is indeed produced by a separate outflow that is tilted more towards the line of sight than the NW/SE outflow (and whose counterflow to the SW is unseen because of obscuration by the disk and envelope). The assumed distance is consistent with the lower end of the range for the IC 348 cluster in the literature (~260 to 320 pc)[37-39]. However, we cannot definitively rule out larger values. A more robust distance constraint awaits detailed 3D modeling of the outflow cavity geometry.

**Radiative Transfer Models**

We calculated models of this protostellar system using a radiative transfer code that treats three separate components: a central protostar, circumstellar disk, and envelope[25]. The envelope is modeled as a collapsing, slowly rotating, spherically symmetric cloud core ("TSC")[40], and the disk as an optically thick, flat disk irradiated by the central star. The disk is a natural outcome of the collapse of a rotating core, with material that is falling in along the equatorial plane landing at the centrifugal radius. In the model calculation, the radiative equilibrium temperature is determined first by using the angle-averaged density distribution in the infall region; at large distances, the envelope is spherically symmetric, but closer to the star the density distribution is flatter since material falls onto the disk. Next, the flux emitted by the system is calculated using the actual flattened, axially symmetric density distribution derived by TSC. Outflow cavities, which follow the streamlines of infalling particles, can also be included in the model. Since there is no material inside the cavity, a large cavity reduces the amount of emitting dust, which for more face-on orientations allows us to see more light from the star and inner disk. In addition, the cavity walls add to scattered light in the near-infrared.

The model parameters that were adjusted to fit observations include the total input luminosity (stellar plus accretion; the latter consists of the accretion luminosity emitted by the disk plus the accretion shock on the stellar surface), the fraction of that luminosity due to the star, the envelope density at a reference radius (which, assuming a mass for the central protostar, can be converted to a mass infall rate[25]), the centrifugal radius (which is also equal to the outer disk radius), the cavity opening angle and the inclination angle. The stellar radius was assumed to be 2 solar radii, the inner disk radius 3 stellar radii, the outer envelope radius 10000 AU. The dust was assumed to be composed of small (<0.3 micron) silicate and graphite grains[41], troilite (FeS) and water ice[42], as well as $CO_2$ ice[43], with abundances similar to the ones adopted in previous models of the well-known protostar L1551 IRS 5[44]. The best-fit parameters were determined by assessing the model fits by eye, after adjusting them based on our extensive previous experience of modeling protostars in the Taurus star-forming region[45]. The near-infrared HST images provided constraints on the inclination and cavity opening angle. The shape of the IRS spectrum in the 15-40 micron region and the 70 micron photometry helped constrain those parameters as well, and more importantly the envelope reference density and centrifugal radius.

Compared to previous model fitting of other protostars, the best-fit models shown in Figure 4 have a relatively small centrifugal radius, typical density, high inclination and large cavity. Lacking direct constraints on the stellar luminosity, we assume it is equal to half the total luminosity in the low flux case; this has at most a small effect on the results since the reddened stellar component is non-negligible only at a narrow range of wavelengths around 3 to 5 microns in the low flux case, and negligible at all wavelengths in the high flux case. Emission from the disk component, attenuated by extinction from envelope dust, dominates the total model flux at 4.5 to 10 microns. The envelope component dominates at wavelengths shorter than 3.6 microns (scattered light) and longer than about 20 microns (dust emission). To match the observed temporal change, we fixed all parameters except for the accretion luminosity, the only parameter for which changes on timescales of weeks are physically plausible.

**Table S1   Observation log**

| instrument | date | reference |
|---|---|---|
| Spitzer/IRAC | 2004 Feb 11 | 46 |
|  | 2004 Sep 8 | 47 |
|  | 2009 Mar 18, 19, 20, 21, 22 | this work |
|  | 2009 Oct 2 - Nov 6 | this work |
|  | 2011 Oct 15 – Nov 22 | this work |
| Spitzer/MIPS | 2004 Feb 21 | 46 |
|  | 2004 Sep 19 | 48 |
|  | 2007 Sep 23, 24, 25, 26, 27 | this work |
|  | 2008 Mar 12, 19 | this work |
| Spitzer/IRS | 2008 Oct 6, 7, 8, 9, 10 | this work |
|  | 2009 Mar 3, 10 | this work |
| HST/WFC3 | 2010 Dec 3, 7, 11, 15, 18, 22, 26 | this work |

All Spitzer observations obtained prior to May 2009 were taken during cryogenic operations; each date corresponds to a separate observation.  The IRAC observations obtained in fall 2009 and fall 2011 were taken during warm operations and only included the first two channels at 3.6 and 4.5 microns; only the range of dates is given, as these consisted of a total of 38 and 20 separate observations, respectively.

**Table S2   Spitzer IRAC photometry**

| MJD (days) | [3.6] | err | [4.5] | err | [5.8] | err | [8] | err |
|---|---|---|---|---|---|---|---|---|
| 53046.020 | 10.830 | 0.005 | 9.563 | 0.004 | 9.070 | 0.005 | 8.967 | 0.006 |
| 53256.453 | 13.190 | 0.025 | 11.854 | 0.028 | 11.239 | 0.034 | 11.088 | 0.049 |
| 54909.242 | 12.767 | 0.044 | 11.509 | 0.038 | 10.977 | 0.045 | 10.871 | 0.055 |
| 54910.121 | 12.771 | 0.041 | 11.517 | 0.035 | 10.932 | 0.047 | 10.855 | 0.053 |
| 54911.043 | 12.607 | 0.034 | 11.368 | 0.027 | 10.762 | 0.034 | 10.628 | 0.045 |
| 54911.938 | 12.535 | 0.024 | 11.262 | 0.024 | 10.698 | 0.035 | 10.572 | 0.044 |
| 54912.863 | 12.385 | 0.023 | 11.084 | 0.020 | 10.530 | 0.027 | 10.388 | 0.028 |
| 55106.898 | 12.682 | 0.015 | 11.379 | 0.035 | | | | |
| 55108.129 | 12.798 | 0.041 | 11.467 | 0.052 | | | | |
| 55109.574 | 12.911 | 0.064 | 11.607 | 0.045 | | | | |
| 55110.496 | 12.993 | 0.063 | 11.685 | 0.044 | | | | |
| 55111.062 | 13.056 | 0.066 | 11.720 | 0.046 | | | | |
| 55112.430 | 13.054 | 0.029 | 11.722 | 0.043 | | | | |
| 55113.066 | 12.967 | 0.045 | 11.614 | 0.038 | | | | |
| 55114.727 | 12.893 | 0.032 | 11.545 | 0.031 | | | | |
| 55115.363 | 12.840 | 0.038 | 11.461 | 0.028 | | | | |
| 55115.961 | 12.685 | 0.029 | 11.349 | 0.025 | | | | |
| 55117.125 | 12.607 | 0.022 | 11.292 | 0.022 | | | | |
| 55118.598 | 12.521 | 0.020 | 11.193 | 0.019 | | | | |
| 55119.246 | 11.943 | 0.014 | 10.605 | 0.012 | | | | |
| 55119.926 | 11.333 | 0.010 | 10.036 | 0.008 | | | | |
| 55107.293 | 12.728 | 0.039 | 11.419 | 0.044 | | | | |
| 55107.570 | 12.756 | 0.037 | 11.406 | 0.052 | | | | |
| 55107.805 | 12.776 | 0.046 | 11.451 | 0.041 | | | | |
| 55108.379 | 12.804 | 0.045 | 11.493 | 0.052 | | | | |
| 55108.574 | 12.820 | 0.044 | 11.526 | 0.049 | | | | |
| 55108.809 | 12.876 | 0.051 | 11.559 | 0.052 | | | | |
| 55116.312 | 12.650 | 0.024 | 11.305 | 0.022 | | | | |
| 55116.613 | 12.654 | 0.021 | 11.261 | 0.023 | | | | |
| 55116.715 | 12.627 | 0.030 | 11.289 | 0.021 | | | | |
| 55117.207 | 12.623 | 0.031 | 11.292 | 0.021 | | | | |
| 55117.457 | 12.581 | 0.027 | 11.277 | 0.017 | | | | |
| 55117.688 | 12.590 | 0.024 | 11.246 | 0.022 | | | | |
| 55123.066 | 10.800 | 0.008 | 9.550 | 0.007 | | | | |
| 55124.953 | 11.023 | 0.011 | 9.804 | 0.009 | | | | |
| 55127.672 | 11.871 | 0.022 | 10.623 | 0.024 | | | | |
| 55129.805 | 12.329 | 0.030 | 11.083 | 0.029 | | | | |
| 55131.270 | 12.482 | 0.036 | 11.224 | 0.036 | | | | |
| 55133.375 | 12.681 | 0.035 | 11.432 | 0.043 | | | | |
| 55135.605 | 12.861 | 0.049 | 11.619 | 0.041 | | | | |

**Table S2 Spitzer IRAC photometry**

| | | | | |
|---|---|---|---|---|
| 55137.516 | 12.934 | 0.050 | 11.663 | 0.037 |
| 55139.059 | 12.675 | 0.025 | 11.382 | 0.027 |
| 55141.508 | 12.708 | 0.033 | 11.415 | 0.028 |
| 55142.109 | 12.407 | 0.020 | 11.127 | 0.022 |
| 55121.203 | 10.732 | 0.006 | 9.493 | 0.005 |
| 55849.770 | 12.134 | 0.018 | 10.856 | 0.020 |
| 55852.410 | 11.452 | 0.014 | 10.194 | 0.010 |
| 55854.035 | 10.782 | 0.008 | 9.544 | 0.007 |
| 55856.422 | 10.212 | 0.006 | 9.022 | 0.005 |
| 55858.234 | 10.515 | 0.009 | 9.343 | 0.007 |
| 55860.555 | 11.113 | 0.017 | 9.909 | 0.016 |
| 55861.828 | 11.578 | 0.025 | 10.374 | 0.020 |
| 55864.434 | 12.251 | 0.042 | 10.991 | 0.040 |
| 55866.211 | 12.459 | 0.041 | 11.173 | 0.038 |
| 55868.066 | 12.582 | 0.040 | 11.297 | 0.043 |
| 55869.809 | 12.717 | 0.048 | 11.461 | 0.040 |
| 55871.445 | 12.928 | 0.055 | 11.633 | 0.042 |
| 55873.859 | 13.020 | 0.039 | 11.728 | 0.033 |
| 55875.852 | 12.461 | 0.020 | 11.149 | 0.018 |
| 55877.824 | 11.903 | 0.011 | 10.665 | 0.010 |
| 55879.949 | 10.623 | 0.006 | 9.451 | 0.006 |
| 55882.020 | 10.210 | 0.005 | 9.094 | 0.005 |
| 55883.836 | 10.505 | 0.007 | 9.369 | 0.007 |
| 55886.422 | 11.132 | 0.017 | 9.971 | 0.016 |
| 55888.137 | 11.692 | 0.025 | 10.488 | 0.028 |

All entries with missing 5.8 and 8 micron photometry correspond to data taken during warm operations.

**Table S3 Spitzer MIPS photometry**

| MJD (days) | [24] | err | [70] | err |
|---|---|---|---|---|
| 53056.062 | 5.79 | 0.05 | -1.07 | 0.18 |
| 53267.371 | 5.70 | 0.05 | -0.92 | 0.21 |
| 53267.594 | 5.42 | 0.06 | | |
| 54366.379 | 5.03 | 0.05 | -1.36 | 0.13 |
| 54367.355 | 5.42 | 0.05 | -1.14 | 0.17 |
| 54368.406 | 5.72 | 0.05 | -0.90 | 0.24 |
| 54369.355 | 5.78 | 0.05 | -0.60 | 0.22 |
| 54370.301 | 5.79 | 0.04 | -0.78 | 0.21 |
| 54538.156 | 3.36 | 0.04 | -1.65 | 0.10 |
| 54544.285 | 4.57 | 0.04 | -1.76 | 0.12 |

The third entry contains no 70 micron data because the scan pattern has a gap at the position of the source.

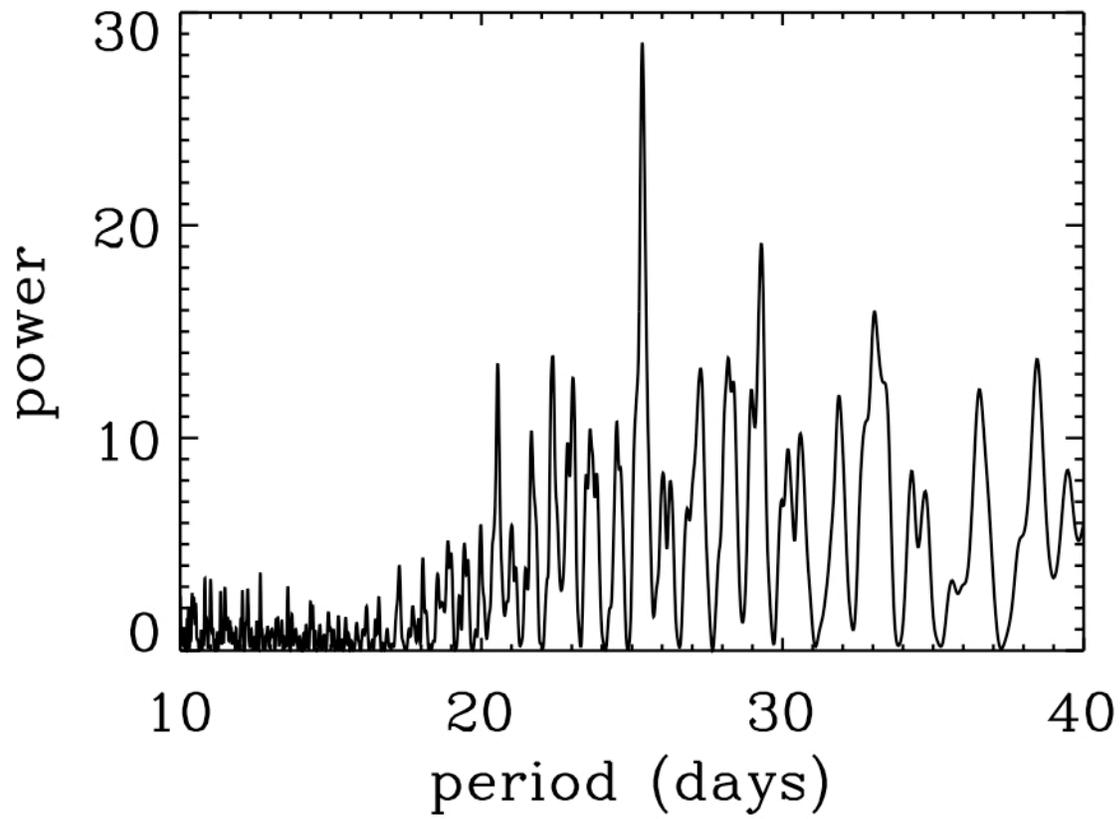

Figure S1. The Lomb-Scargle periodogram corresponding to the Spitzer photometry of LRLL 54361. The peak power corresponds to a period of 25.34 days.

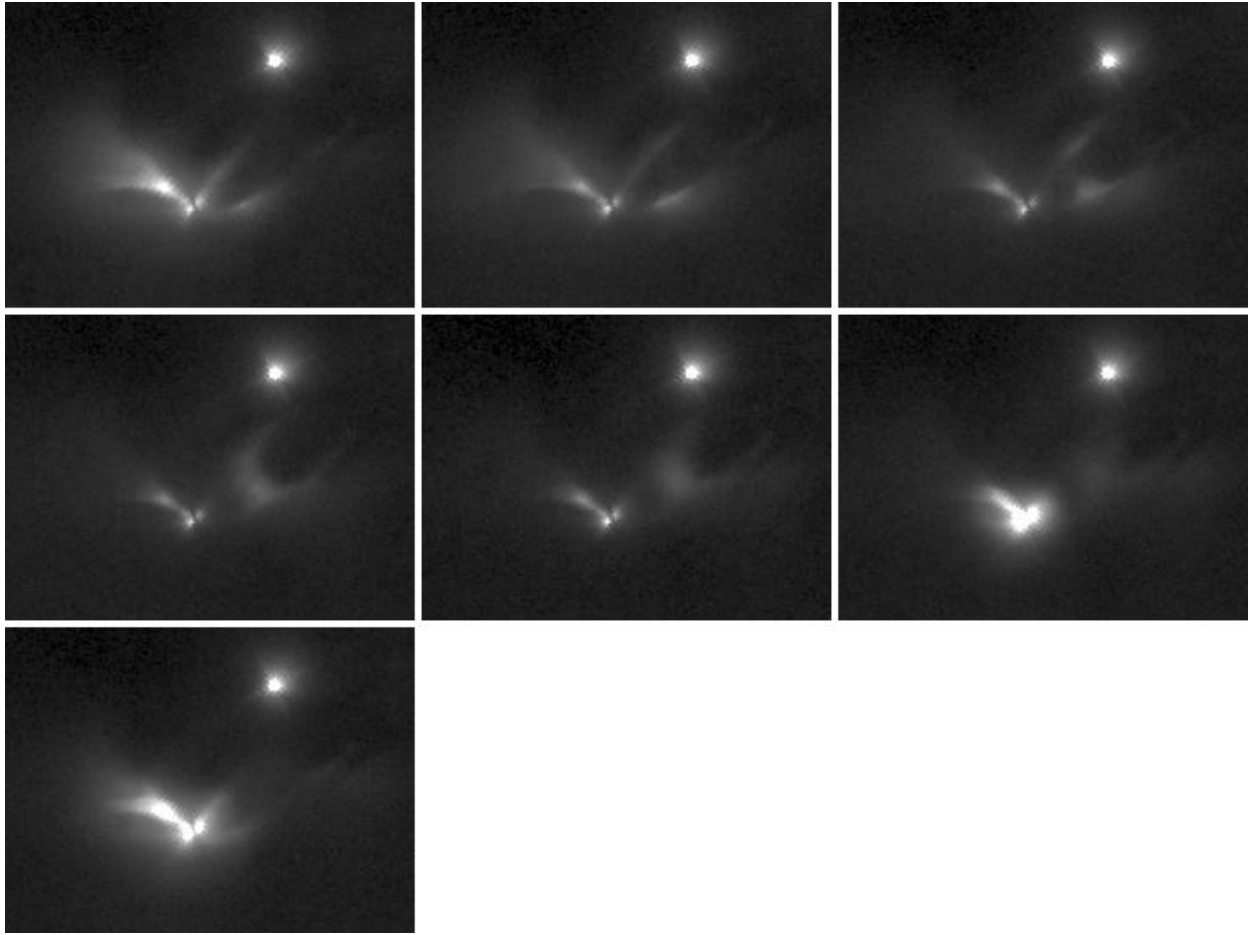

Figure S2. A montage of the HST/WFC3 F160W images, centered roughly on L54361 (only a portion of the full images are shown). North is up, east is to the left. Each panel spans about 25"x20". The observations are separated by about 3 - 4 days (see Table 1); time proceeds from left to right, top to bottom. Another cluster member, LRLL 1843, is seen at upper right.

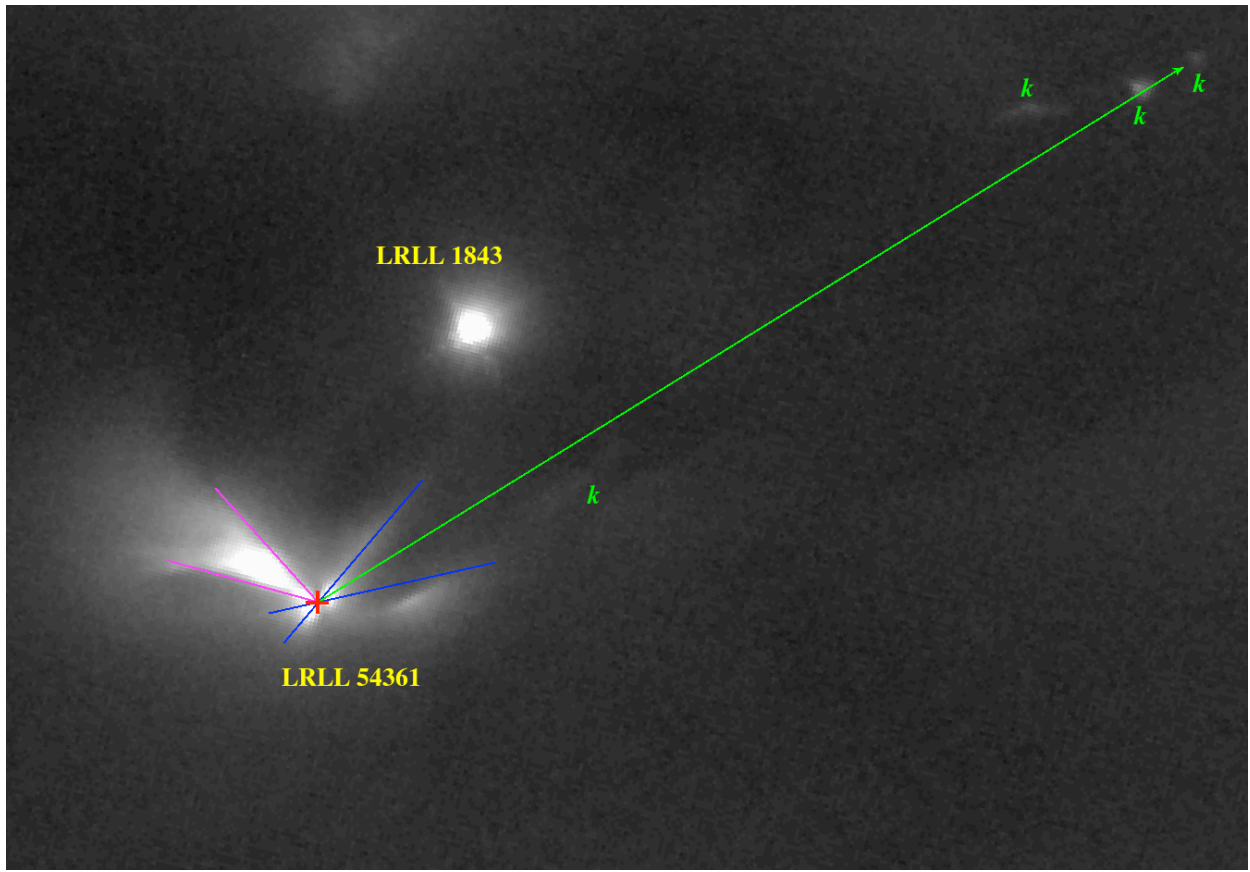

Figure S3. Geometry of the L54361 system. The grayscale image is from the HST/WFC3 epoch 1 image. The red cross marks the centroid position of the Spitzer/IRAC source. The blue lines mark the approximate boundaries of the NW/SE outflow cavities. The magenta lines mark the rough boundaries of a separate scattered light region that may be a cavity produced by a second misaligned outflow, or possibly an asymmetric extension of the SE outflow cavity. The green arrow indicates the general direction of the NW jet, with individual knots marked with a "k" (note that the positions and morphologies of these do not change with time). The brown dwarf cluster member LRLL 1843 is also labeled.

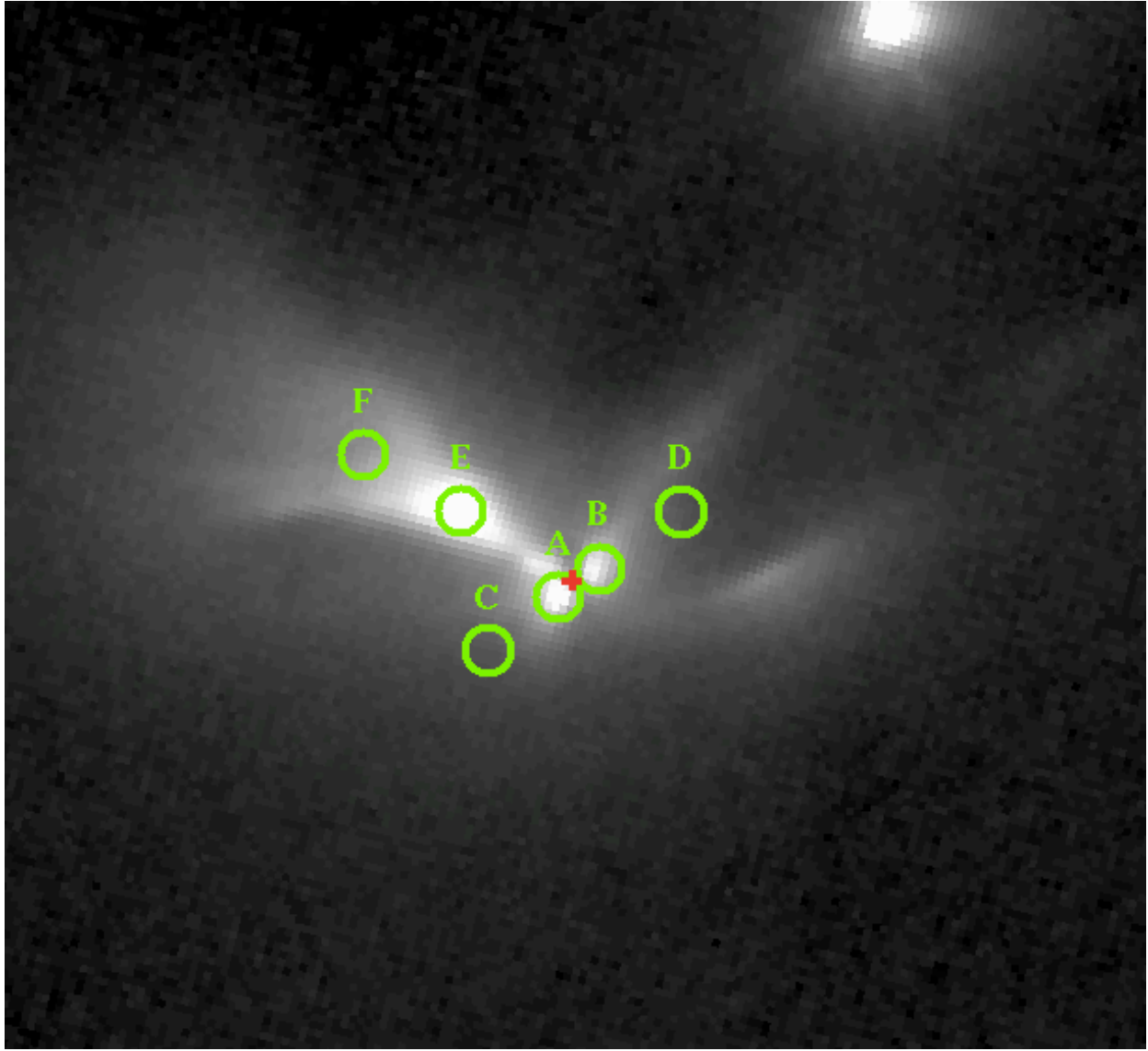

Figure S4. Close-up of the epoch 1 HST/WFC3 image. The red cross marks the centroid position of the Spitzer source. The green circles mark the six regions of aperture photometry shown in Figure S5.

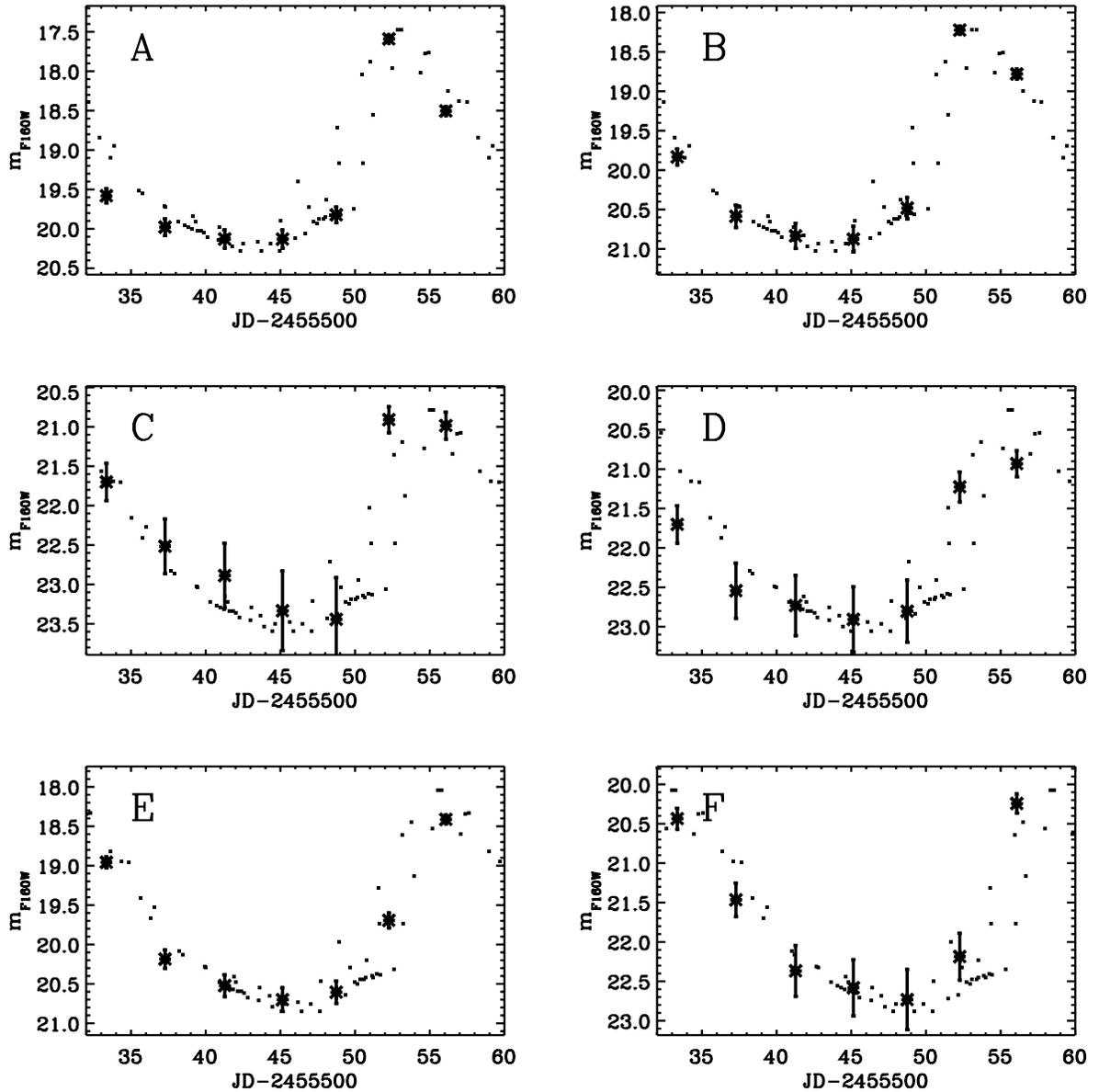

Figure S5. Aperture photometry of six regions from HST/WFC-IR F160W images of LRLL 54361 (asterisks with error bars) as a function of observation time. Each panel shows the aperture magnitudes corresponding to the labeled region shown in Figure S4. The small squares show all warm Spitzer/IRAC 3.6 micron magnitudes, offset to match the F160W magnitude range. The Spitzer epochs were phased by the period of 25.34 days and then shifted with respect to the HST observing times by the following lags, calculated as described in the text: 0.51 days (A), 0.75 days (B), 2.65 days (C), 3.16 days (D), 3.22 days (E), 5.98 days (F).